\documentclass[aps,prl,twocolumn,floatfix, superscriptaddress]{revtex4}
\usepackage[dvipdfx,cmyk,natbib]{xcolor}     % use if color is used in text
\usepackage{amsmath}
\usepackage{amsfonts}
\usepackage[caption=false]{subfig}
\usepackage{graphicx}

\renewcommand{\vec}[1]{\mathbf{#1}}

\newcommand{\mat}[1]{\mathrm{#1}}
\newcommand{\by}{\times}
\newcommand{\of}[1]{\!\left(#1\right)}
\newcommand{\pdf}{{pdf}}
\newcommand{\abs}[1]{\left|#1\right|}
\newcommand{\prob}[1]{\mathcal{#1}}

\newcommand{\ket}[1]{\left|#1\right\rangle}

\begin{document}

\title{Direct dialling of Haar random unitary matrices}
\author{Nicholas J. Russell}
\affiliation{Centre for Quantum Photonics, H. H. Wills Physics Laboratory \& Department of Electrical and Electronic Engineering, University of Bristol, BS8 1UB, UK}
\author{Levon Chakhmakhchyan}
\email{levon.chakhmakhchyan@ulb.ac.be}
\affiliation{Centre for Quantum Information and Communication, Ecole polytechnique de Bruxelles, CP 165, Universit\'{e} libre de Bruxelles, 1050 Brussels, Belgium}
\author{Jeremy L. O'Brien}
\affiliation{Centre for Quantum Photonics, H. H. Wills Physics Laboratory \& Department of Electrical and Electronic Engineering, University of Bristol, BS8 1UB, UK}
\author{Anthony Laing}
\email{anthony.laing@bristol.ac.uk}
\affiliation{Centre for Quantum Photonics, H. H. Wills Physics Laboratory \& Department of Electrical and Electronic Engineering, University of Bristol, BS8 1UB, UK}

\date{\today}

\begin{abstract}
Random unitary matrices find a number of applications in quantum information science, and are central to the recently defined boson sampling algorithm for photons in linear optics.
We describe an operationally simple method to directly implement Haar random unitary matrices in optical circuits, with no requirement for prior or explicit matrix calculations.
Our physically-motivated and compact representation directly maps independent probability density functions for parameters in Haar random unitary matrices, to optical circuit components.
We go on to extend the results to the case of random unitaries for qubits.
\end{abstract}

\maketitle

\begin{figure}[t]
  \centering
  \subfloat[]{
    \includegraphics{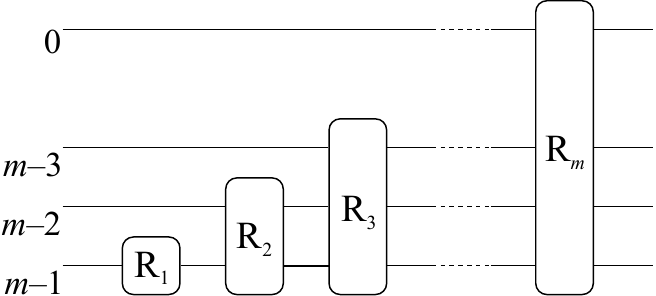}
    \label{fig:recursive}
  } \\
  \subfloat[]{
    \includegraphics{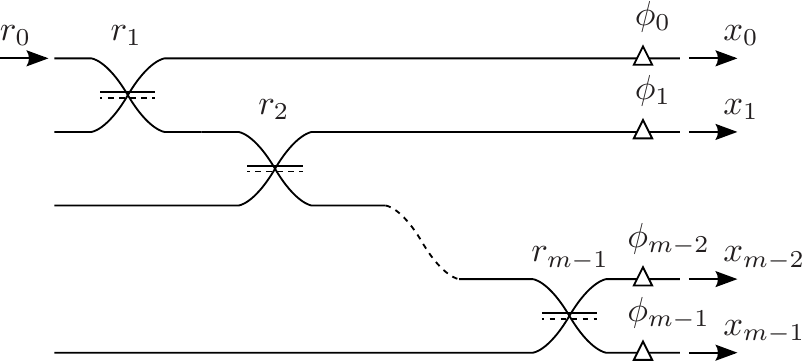}
    \label{fig:cascade}
  }
  \caption{
Recursive decomposition of a unitary in the triangular scheme.
(a) An \(m \by m\) unitary transformation can be factored as a product of \(m\) unitary transformations \(\mat{R}_{i}\), each acting on a successively larger subspace (the subscripts 0, ..., $m-1$ on the left label the modes of the transformation).
(b) A linear optical \(\mat{R}_{m}\) can be constructed from a cascade of beamsplitters and phase shifts on optical modes.
Both the \emph{Cartesian} basis, \(\vec{x}\) and the \emph{physical} basis, \(\vec{r}\) are illustrated.}
  \label{fig:reck}
\end{figure}

The development of the boson sampling problem \cite{aa-conf-11-333, cr-nat-7-545, br-sci-339-794, sp-sci-339-798, ti-nphoton-7-540} has motivated fresh interest in studying Haar random unitary matrices (HRUs) \cite{zy-jpa-27-4235} realised with optical circuits to act on multiphoton states. Simultaneously, developments in integrated optics
\cite{Politi:2008tl,Obrien:2009eu,Marshall:2009ub,Crespi:2011cy,Matthews:2009gi,Smith:2009hm,Laing:2010fk,Peruzzo:2010tq,Sansoni:2012eq}
now facilitate the construction of large-scale optical circuits capable of actively realising any unitary operator \cite{ulo} including HRUs. Furthermore, HRUs play an important role in various tasks for quantum cryptography \cite{Hayden:2004} and quantum information protocols \cite{Bennett:2005, Abeyesinghe:2009}, as well as the construction of algorithms \cite{Sen:2006}.

Here we present a simple procedure for choosing a HRU on an optical circuit, implemented in terms of recursive decompositions of a unitary operator~\cite{re-prl-73-581, oxford}, by choosing values of the physical parameters independently from simple distributions. This procedure is useful for applications where the exact unitary description of the implemented circuit is less important than a guarantee that it is drawn from the correct distribution.
While similar parameterisations exist in the mathematical literature \cite{sp-jmp-53-013501}, an operational application within linear optics is not widely appreciated. We extend the result to systems of qubits, by deriving a mapping between a linear-optical circuit on \(m=2^{n}\) modes and a circuit operating on \(n\) qubits. Note that constructions for pseudo-HRUs on qudit and qubit systems are also available, serving as a general framework to investigate randomising operations in complex quantum many-body systems \cite{Nakamata:2016, Brandao:2012}.

\begin{figure*}[t]
  \includegraphics[width=.85\linewidth]{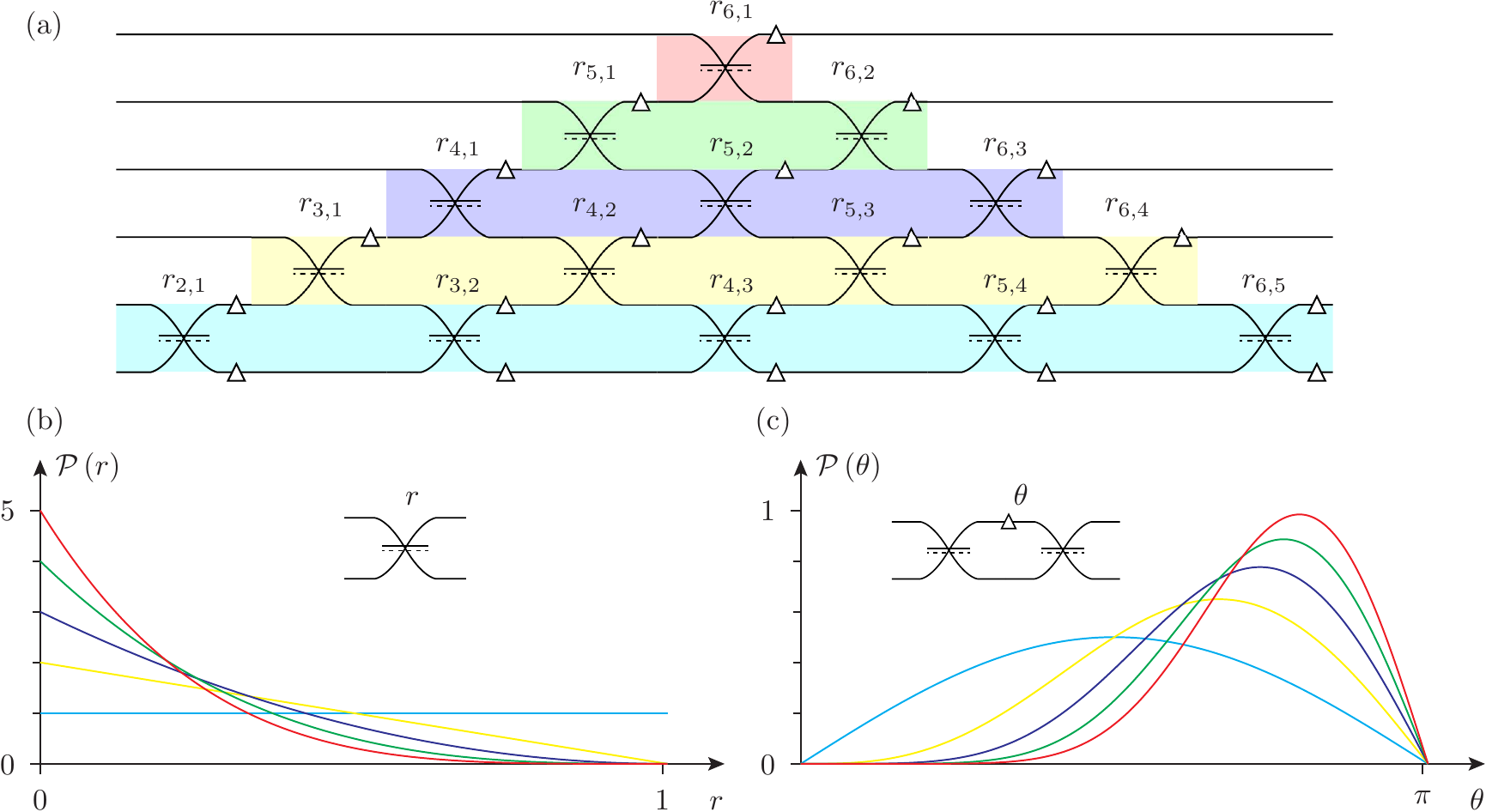}
  \caption{
Direct dialling of a HRU in a triangular linear optical circuit.
(a) A $6 \by 6$ unitary operator implemented with a triangular six mode linear optical circuit.
(b) The \pdf{}s from which beamsplitter reflectivities should be chosen to directly implement a HRU.  Those on higher rows are chosen according to polynomials with increasing bias towards lower reflectivities.
(c) A variable reflectivity beamsplitter can be effectively implemented with an MZI composed of a phase shift, \(\theta\) between two $1/2$ reflectivity beamsplitters. The MZI phases may be chosen directly from the distributions shown, with increasing bias towards \(\theta=\pi\). The integrated optics implementation with directional couplers results in a bias towards \(\theta=0\).
Line colours in (b) and (c) correspond to shading in (a).}
  \label{fig:example}
\end{figure*}

Choosing a HRU is analogous to choosing a random number from a uniform distribution, in that it should be unbiased. The probability of selecting a particular unitary matrix from some region in the space of all unitary matrices should be in direct proportion to the volume of the region as defined by the Haar measure, which is the unique translation-invariant measure on the space of unitary matrices.
As argued in reference~\cite{re-phd},
the columns of an $m$-dimensional HRU may be made up from vectors \(\left\{v_{i} \right\} = \left\{ v_{1}, v_{2}, \cdots, v_{m} \right\}\) that are successively drawn from the unbiased distribution of unit vectors in the subspace of \(\left(m-i+1\right)\) dimensions, orthogonal to all previous vectors. The problem of choosing HRUs thus reduces to the problem of recursively choosing such a set of orthogonal vectors.

As we will show, this approach is particularly relevant to recursive circuit decompositions that allow any unitary matrix to be implemented over $m$ optical modes, by choosing appropriate values for beamsplitter reflectivities and phase shifters.
We first consider the \emph{triangular scheme} \cite{ulo} shown in figures~\ref{fig:reck},~\ref{fig:example}a, which is a variant of that proposed by Reck et al.~\cite{re-prl-73-581}, and which represents an $m\times m$ unitary matrix $\mathrm{U}$ as a product of unitary operators labeled $\mathrm{R}_n$, $\mathrm{U}=\prod_{i=0}^{m-1}\mathrm{R}_{m-i}$
\footnote{
Note that scheme realised in the integrated photonic chip of reference~\cite{ulo},
in which each beamsplitter in every block R$_n$ couples two adjacent modes,
differs from the earlier proposal of reference~\cite{re-prl-73-581},
in which the first mode is consecutively coupled with modes  $2, 3, ..., n$.
}.
Each block \(\mat{R}_{n}\) is chosen to transform the mode $j=m-n$, or the corresponding basis state, denoted by $\ket{\Psi_{m-n}}$, into the \(n\)-dimensional unit vector \(\ket{v_{n}}\), over modes $j=m-n$ to $k=m-1$, i.e.,
\begin{equation}\label{vn}
\ket{v_n}=\mathrm{R}_n \ket{\Psi_{m-n}}.
\end{equation}
This vector undergoes further transformations under subsequent blocks \(\mat{R}_{i}\) ($n<i\leq m$) to finally produce \(\ket{f_{n}}\) that occupies all $m$ modes:
\begin{equation}
\ket{f_n}=\mathrm{R}_m\cdots\mathrm{R}_{n+1} \ket{v_n}.
\end{equation}
Orthogonality between each of the $m$ \(\ket{f_{i}}\) vectors is guaranteed.
Further, if the vector \(\ket{v_{n}}\) is chosen from the unbiased distribution of unit vectors in \(n\) dimensions, the property of left invariance ensures that \(\ket{f_{n}}\) does not become biased by the operation of the subsequent \(\mat{R}_{i}\).

The next and main task is therefore determining how an unbiased vector in \(n\) dimensions may be implemented with \(\mat{R}_{n}\) by choosing values for the linear optical components from which it is constructed, according to the expansion shown in figure~\ref{fig:reck}(b). To achieve this, consider the complex Gaussian vector in \(n\) dimensions:
\begin{equation}\label{vn2}
\ket{\vec{v}_{n}} = \sum_{i=0}^{n-1} z_{i} \ket{\Psi_i} = \sum_{i=0}^{n-1} \tau_{i}
e^{i \alpha_{i}} \ket{\Psi_i},
\end{equation}
%
%\begin{equation}\label{vn2}
 % \ket{\vec{v}_{n}} = \sum_{i=0}^{n-1} z_{i} \ket{i} = \sum_{i=0}^{n-1} \tau_{i}
 %e^{i \alpha_{i}} \ket{i},
%\end{equation}
%
where $\ket{\Psi_i}$ denotes the $i$th basis state and the \(z_i\) are independent and identically distributed normal random variables with the  probability density function (pdf), \( \prob{P}_{z_i} \of{z} =1/\pi \exp \left( -\abs{z}^2 \right) \).
This independence means that the \pdf{} for \(\vec{v}_n\) is the product of the \pdf{}s for these elements and depends only on the magnitude of the vector:
\begin{equation}
  \label{eq:vec}
  \prob{P}_{\vec{v}_n} (\vec{x}) =\frac{1}{\pi^n} e^{ -\left(x_0 + x_1 + \dots + x_{n-1} \right)} = \frac{1}{\pi^n} e^{-\abs{\vec{v}_n}^2},
\end{equation}
where \(x_i = \abs{z_i}^2 \).
%
%Now consider the change of variables from this basis~\(\vec{x}\), which we call the Cartesian basis, to a new basis,~\(\vec{r}\):
We now show how this basis~\(\vec{x}\), which we call the Cartesian basis, can be mapped to a new basis,~\(\vec{r}\). We call the latter the physical basis, since, as we demonstrate below, it contains the variables corresponding directly to components in a physical realisation of the vector in linear optics. Namely, we denote by \(r_0\) the power of the input to the given block R$_n$, while the other \(r_i\) stand for the reflectivities of beamsplitters (see also figure~\ref{fig:reck}(b)). Next, combining the definition (\ref{vn}) and equation (\ref{vn2}), we find
\begin{eqnarray}\label{eq:map:11}
&&z_0=e^{i \phi_0}\sqrt{r_0 r_1}\\
&&z_i= e^{i \phi_i}\sqrt{r_0r_{i+1}}\prod_{k=1}^{i} \sqrt{1-r_k} \,\,\,\,\,\,\,\, 0 < i \leq n-1,
\end{eqnarray}
where the matrix (in the Pauli basis) $B(r)=\sqrt{r} \sigma_{z} + \sqrt{1-r} \sigma_{x}$ has been used to describe a beamsplitter as a function of its reflectivity. Finally, taking into account that $x_i=|z_i|^2$, we find,
\begin{align}
  \label{eq:map}
  r_0 &= \sum_{k=0}^{n-1} x_k \\
    \label{eq:map1}
  r_i &= \frac{x_{i-1}}{\sum_{k=i-1}^{n-1} x_k} & 0 < i \leq n-1  \\
    \label{eq:map2}
  \phi_i &= \alpha_{i}.
\end{align}
%
%We refer to \(\vec{r}\) as the physical basis because the variables correspond directly to components in a physical %realisation of the vector in linear optics. In particular, \(r_0\) is analogous to the power of the input while the other \(r_i\) are %reflectivities of beamsplitters. The relationship between \(\vec{x}\), \(\vec{r}\) and the physical system is shown in %figure~\ref{fig:reck}.

We must show that the \pdf{}s for the vector \(\vec{v}_n\) are separable in the physical basis so that the experimental parameters can be chosen independently.
We also need to derive the form of the marginal distributions for the \(r_i\) and \(\phi_{i}\), from which experimental parameters must be chosen to obtain a Haar unitary.
Since there is no functional dependence on the \(\alpha_{i}\) parameters in equation~(\ref{eq:vec}) and there is a one-to-one mapping \(\alpha_{i} \rightarrow \phi_{i}\), these phases can be chosen uniformly and independently from the interval \(\left[0,2\pi\right)\).

Finding the \pdf{}s for the beamsplitter reflectivities requires a more careful change in bases, using the Jacobian,
\begin{equation}
  \prob{P}_{\vec{v}_n} \of{\vec{r}} = \prob{P}_{\vec{v}_n} \of{\vec{x}}
  \abs{\det \mat{J} \of{\vec{x}, \vec{r}}}.
\end{equation}
The pre-factor from (\ref{eq:vec}) is expressed in the \(\vec{r}\) basis simply as \(\exp
\of{-r_0}\), so is trivially separable.
We therefore consider the Jacobian matrix
\begin{equation}
  \mat{J}_{i,j} \of{\vec{x}, \vec{r}} = \frac{\partial x_{i}}{\partial r_j}
\end{equation}
with
\begin{align}
x_0& =r_0\; r_1\\
x_i & = r_{0} \; r_{i+1} \prod_{k=1}^{i} ( 1-r_{k}) & 0 < i \leq n-1 .
\label{eqXToR}
\end{align}
For the four cases
\begin{align}
  \label{eq:firstcolumn}
  \mat{J}_{i,j} &= r_{i+1} \prod_{k=1}^{i} \left( 1-r_k \right) && j=0 \\
  \mat{J}_{i,j} &= \frac{-r_0 r_{i+1}}{1-r_{j}} \prod_{k=1}^{i} \left( 1-r_k
    \right) && 0< j \leq i\\
  \label{eq:diagonal}
  \mat{J}_{i,j} &= r_0 \prod_{k=1}^{i} \left( 1-r_k \right) && j = i+1 \\
  \mat{J}_{i,j} &= 0 && j>i+1
\end{align}
where the variable \(r_{n}=1\) has been introduced for convenience.

We show that this form of matrix (lower Hessenberg) can always be transformed into a lower triangular matrix---for which the determinant is simply the product of the diagonal elements---by elementary operations, which do not change the absolute value of the determinant.

The first step is to perform a set of operations on the \(j=0\) column, \(\vec{c}_{0}\), that set the upper \(n-1\) terms to zero, as follows:
\begin{equation}
  \vec{c}_{0}^{\left( k \right)} = \vec{c}_{0}^{\left( k-1 \right)} -
  \vec{c}_{k} \frac{\mat{J}_{k-1,0}^{\left( k-1 \right)}}{\mat{J}_{k-1,k}},
\end{equation}
where \(k\) runs from \(1\) to \(m-1\),
$\vec{c}_{0}^{(k)}$ and $\mat{J}_{k,0}^{\left( k \right)}$ are those quantities after $k$ operations,
and
$\vec{c}_{k}$ is the $k$th column.
We can then place the column \(\vec{c}_{0}\) as the rightmost column, at which point the matrix is lower triangular.
 After the procedure is complete the element \(\mat{J}_{0,n-1}=1\) (see appendix for detailed proof).

The Jacobian determinant is given by multiplying the diagonal elements of the shifted matrix
\begin{equation}
  \det \mat{J} \of{ \vec{x}, \vec{r} } = \prod_{i=1}^{n-1} \mat{J^{\prime}}_{i,i}
\end{equation}
which are given by equation~(\ref{eq:diagonal}).
The explicit form of the \pdf{} in the \(\vec{r}\) basis is
\begin{equation}
  \prob{P}_{\vec{v}_n} \of{ \vec{r} } = e^{-r_0} r_0^{n-1} \prod_{k=1}^{n-1}
  \left( 1-r_k \right)^{n-k-1},
\end{equation}
which is manifestly separable.

It can be verified by explicit integration that this expression is appropriately
normalised. Since the \pdf{} is separable in this basis, the variables \( r_i
\) are independent, and can be chosen according to their marginal distributions,
\begin{align}
\label{eq:pdf}
  \prob{P}_{r_{n,i}} \of{r} = \left( n-i \right) \left( 1-r \right)^{n-i-1} &&
  1 \leq i < n,
\end{align}
where, for clarity, \( r_{n,i} \) denotes the reflectivity of the $i$th beamsplitter
in the $n$th rotation, \( \mat{R}_n \).
We now integrate over \(r_{0}\) to obtain a compact form for the \pdf{} of
\(n\)-dimensional \emph{unit} vectors,
\begin{equation}
  \prob{P}_{ v_n } (\vec{r}) = \left( n-1 \right)! \prod_{k=1}^{n-1} \left( 1-r_{n,k}
  \right)^{n-k-1}
\end{equation}
and express the \pdf{} for the full circuit of beamsplitters,
\(\prob{P}_{\vec{C}}(\vec{r})\) as the product of the \pdf{}s for the diagonal arrays of
beamsplitters:
\begin{equation}
  \prob{P}_{\vec{C}}(\vec{r}) = \prod_{j=1}^{m} \left[ \left( j-1 \right)!
  \prod_{k=1}^{j-1} \left( 1-r_{j,k} \right)^{j-k-1} \right].
\end{equation}

%Here, the matrix (in the Pauli basis) $B(r)=\sqrt{r} \sigma_{z} + \sqrt{1-r} \sigma_{x}$ has been used to describe a %beamsplitter as a function of its reflectivity.
%
Recalling the beamsplitter transformation $B(r)=\sqrt{r} \sigma_{z} + \sqrt{1-r} \sigma_{x}$, we note that a variable reflectivity beamsplitter can be constructed as a Mach-Zehnder interferometer (MZI),
from a variable phase shifter $\theta$ between two $1/2$ reflectivity beamsplitters, $H=B(1/2)$,
to give $B_{v}(\theta)=\cos \frac{\theta}{2}I + i \sin \frac{\theta}{2} \sigma_{x}$ (up to a global phase).
It is then useful to re-express the \pdf{}s in terms of MZI phase shifts.
The further change of variables, $r = \cos^{2} \frac{\theta}{2}$, gives
\begin{equation}
  \prob{P}^\mathrm{B}_{\theta_{i}} \of{ \theta }
  =
  \left( n-i \right)\cos\frac{\theta}{2}\left[\sin\frac{\theta}{2}\right]^{2 \left(n-i\right) -1}.
\label{eq:pdf:phase}
\end{equation}
In the setting of integrated optics, where beamsplitters are implemented with directional couplers on waveguides according to
$D(1/2)=\frac{1}{\sqrt{2}} ( I + i \sigma_{x} ) $ for reflectivity of $1/2$,
the \pdf{}s are given by (\ref{eq:pdf:phase})
but with $\sin$ and $\cos$ functions interchanged, i.e.,
\begin{equation}
\prob{P}^\mathrm{D}_{\theta_{i}} \of{ \theta }
=
\left( n-i \right)\sin\frac{\theta}{2}\left[\cos\frac{\theta}{2}\right]^{2 \left(n-i\right) -1}.
\label{eq:pdf:phase:D}
\end{equation}

In practical terms, an optical circuit composed of beamsplitters and variable phase shifters can directly \emph{dial up} a configuration corresponding to a HRU, by choosing phase shifter values from the derived \pdf{}s. A six mode example is given in figure~\ref{fig:example}.

We note that the version of the triangular scheme used here, in which each beamsplitter in every block R$_n$ couples two adjacent modes, differs from the original scheme~\cite{re-prl-73-581}, in which the first mode is consecutively coupled with modes  $2, 3, ..., n$. It is easy to check, however, that the mapping (\ref{eq:map})-(\ref{eq:map2}) can be applied to the original scheme as well, by replacing $r$ with $1-r$ and relabelling the output modes: $\{x_0, x_1, ..., x_{m-1}\}\rightarrow \{x_{m-1}, x_0, ..., x_{m-2}\} $. Such a change of variables does not affect the Jacobian determinant and the final expression for the reflectivity pdfs for the original scheme is obtained by replacing $r$ with $1-r$ in equation~(\ref{eq:pdf}) (the phases are again chosen uniformly and independently from the interval $[0, 2 \pi)$).

\begin{figure}[t]
	\includegraphics[trim=75 0 110 0, clip, width=1\linewidth]{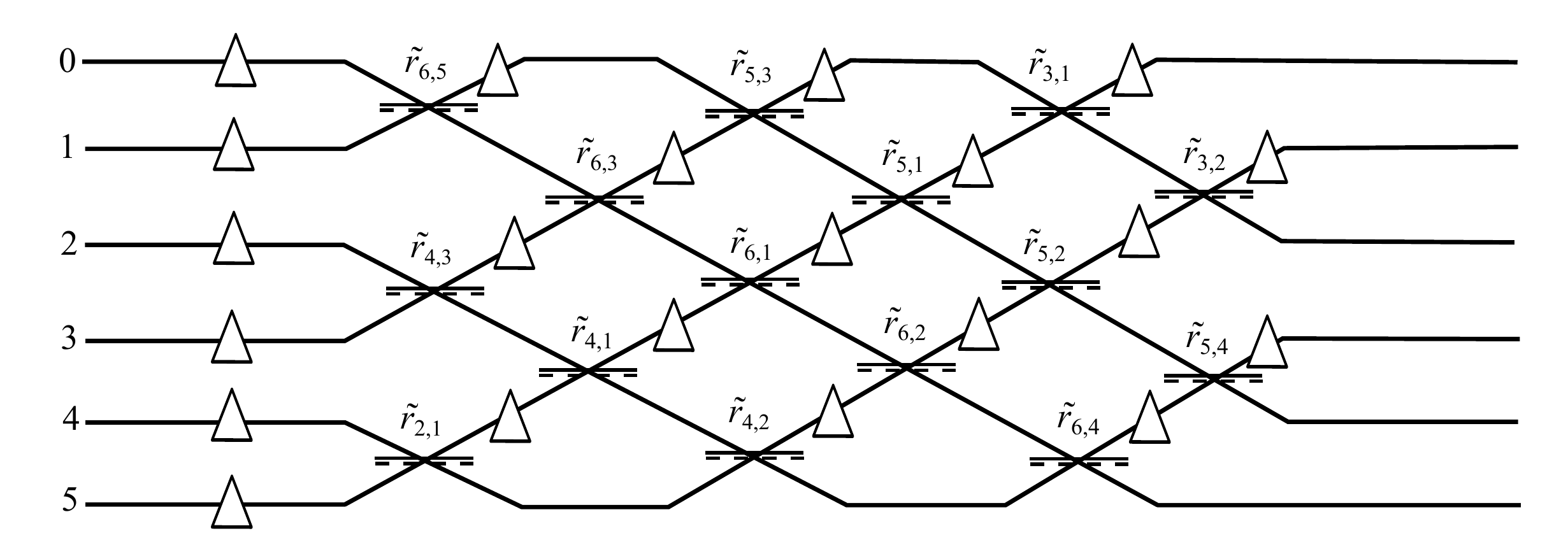}
	\caption{
	A $6\times 6$ unitary operator implemented with a six-mode linear optical circuit according to the rectangular scheme. Here, $\tilde{r}_{n,i}$ stands for the reflectivity of the $i$th beamsplitter of the block $\tilde{R}_n$. Within each $\tilde{R}_n$ ($n=2, ..., 6$), we enumerate the beamsplitters according to the sequence $s$, which consists of $n-1$ indices, with odd (even) numbers arranged in descending order and followed by even (odd) numbers, arranged in ascending order (see also main text). In this figure, the beamsplitters in the $i$th row mix the modes $i-1$ and $i$ (e.g., the beamsplitters of the third row, $\tilde{r}_{4,3}$, $\tilde{r}_{6,1}$ and $\tilde{r}_{5,2}$, couple the modes 2 and 3).}
	\label{fig:oxford}
\end{figure}

Next, we analyse the alternative decomposition of unitary matrices, proposed by Clements et al.~\cite{oxford}, which corresponds to a {\it rectangular} mesh of beamsplitters and phase shifters, as shown in figure~\ref{fig:oxford} for six modes.
While the triangular scheme might be more resilient to loss and other errors in experiments in which only a small proportion of its (upper) input ports are accessed,
the rectangular scheme is likely to be beneficial for experiments that involve accessing most of its inputs. The more compact rectangular scheme may also fit a greater number of modes on standard wafers used in the fabrication of integrated photonic circuits.

The rectangular scheme obeys the blocked structure, analogous to the triangular scheme described above. That is, an $m\times m$ unitary matrix $\mathrm{U}$ can be written down as a product of blocks $\tilde{\mathrm{R}}_n$ (hereafter the tilde refers to the decomposition of reference~\cite{oxford}). Each of these blocks, as previously, transforms the mode $m-n$ into a vector over modes $m-n$ up to $m-1$ (see also figure~\ref{fig:reck}(b)). More precisely, for odd (even) $m$, $\mathrm{U}=\prod_{j=1}^{m/2}\tilde{\mathrm{R}}_{2j-1}\prod_{i=0}^{m/2-1}\tilde{\mathrm{R}}_{m-2i}$ ($\mathrm{U}=\prod_{j=1}^{(m-1)/2}\tilde{\mathrm{R}}_{2j}\prod_{i=0}^{(m-1)/2}\tilde{\mathrm{R}}_{m-2i}$). Moreover, the mapping of equations~(\ref{eq:map})-(\ref{eq:map2}) for the operator $\mathrm{R}_n$ for the triangular scheme can be used for $\tilde{\mathrm{R}}_n$ as well, by a simple substitution. Namely, for even (odd) $m$ we replace  $\tilde{r}_{n,i}$ by $r_{n,s(i)}$, $\forall{i}$, where $s$ is a sequence of $n-1$ indices, with odd (even) numbers arranged in descending order and followed by even (odd) numbers, arranged in ascending order (e.g., for $m=n=6$, we have $s=\{5, 3, 1, 2, 4\}$).

This substitution leaves the corresponding Jacobian determinant unaffected. Therefore, the pdfs given in equation~(\ref{eq:pdf}) for reflectivities $r_{n,i}$ for the triangular scheme correspond to that of the rectangular scheme, but for $\tilde{r}_{n,s(i)}$. In other words, $\mathcal{P}_{r_{n,i}}(r)=\tilde{\mathcal{P}}_{\tilde{r}_{n,s(i)}}(r)$. Subsequently, we find
\begin{equation}
\tilde{\mathcal{P}}_{\tilde{r}_{n,i}}(\tilde{r})=[n-s(i)](1-\tilde{r})^{n-s(i)-1}.
\end{equation}
Alternatively, one can reorder the reflectivities $\tilde{r}_{n,i}$ according to the sequence $s$, as is done in figure~\ref{fig:oxford}, yielding $\mathcal{P}_{r_{n,i}}(r)=\tilde{\mathcal{P}}_{\tilde{r}_{n,i}}(r)$. Finally, the phases of the rectangular scheme, analogous to the triangular scheme, are chosen uniformly and independently from the interval $[0, 2\pi)$.

\begin{figure}[t]
	\includegraphics[width=0.45\textwidth]{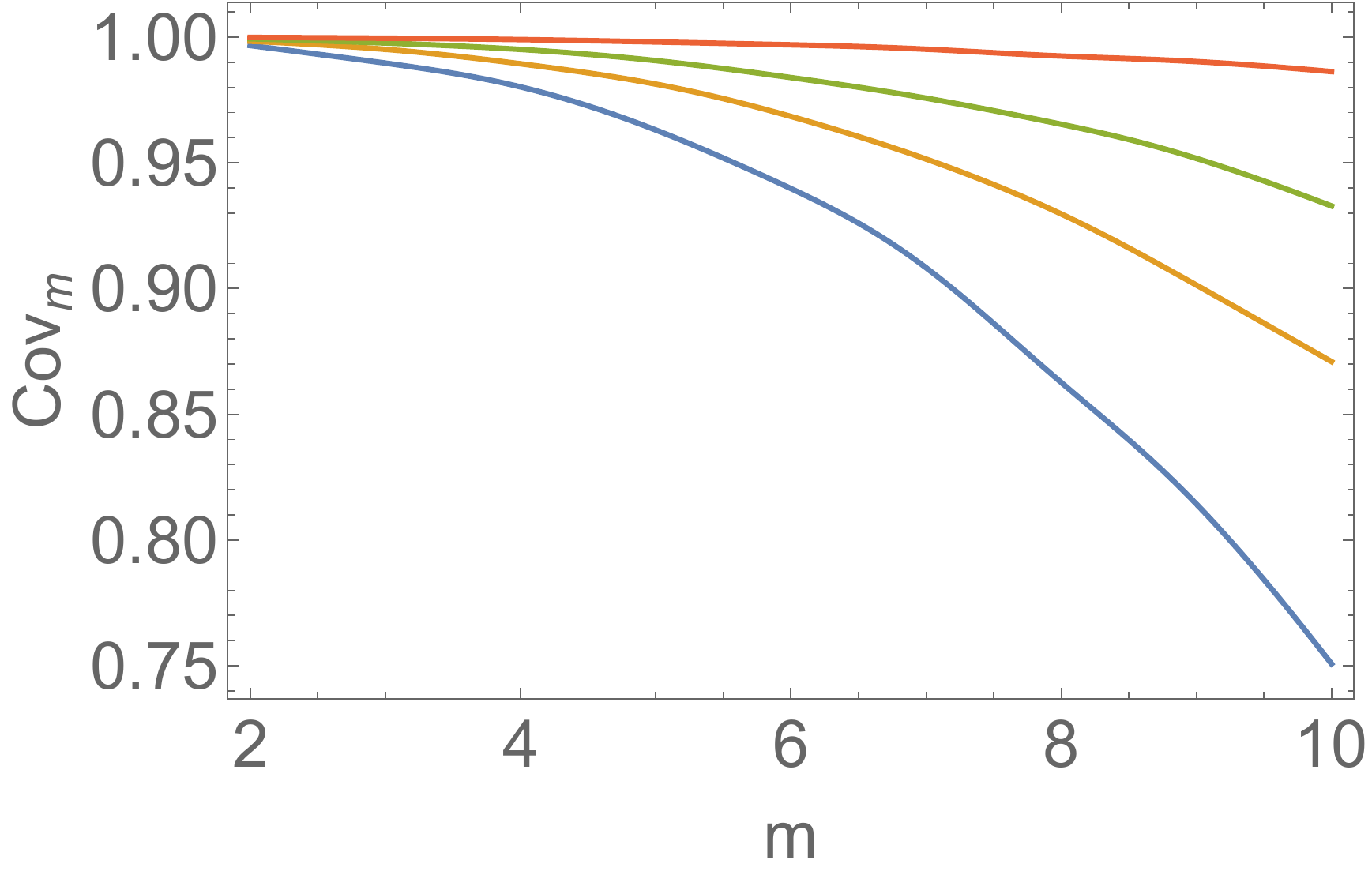}
	\caption{
		Coverage $\mathrm{cov}_m$ of the unitary space versus the circuit size $m$. The phase shifters are assumed to cover their full range $[0, 2 \pi)$, while the range of reflectivities is restricted to $[|\varepsilon|, 1-|\varepsilon|]$, where random errors $\varepsilon$ are drawn from a zero-mean normal distribution. The curves correspond to different variances $\sigma$ of the errors ($\sigma=\{1, 5, 10, 20\} \times 10^{-4}$ from the upper to the bottom curve). For each $m$, the coverage is averaged over many realizations of $\varepsilon$.}
	\label{fig:coverage}
\end{figure}

\begin{figure*}[t]
  \includegraphics[width=.9\linewidth]{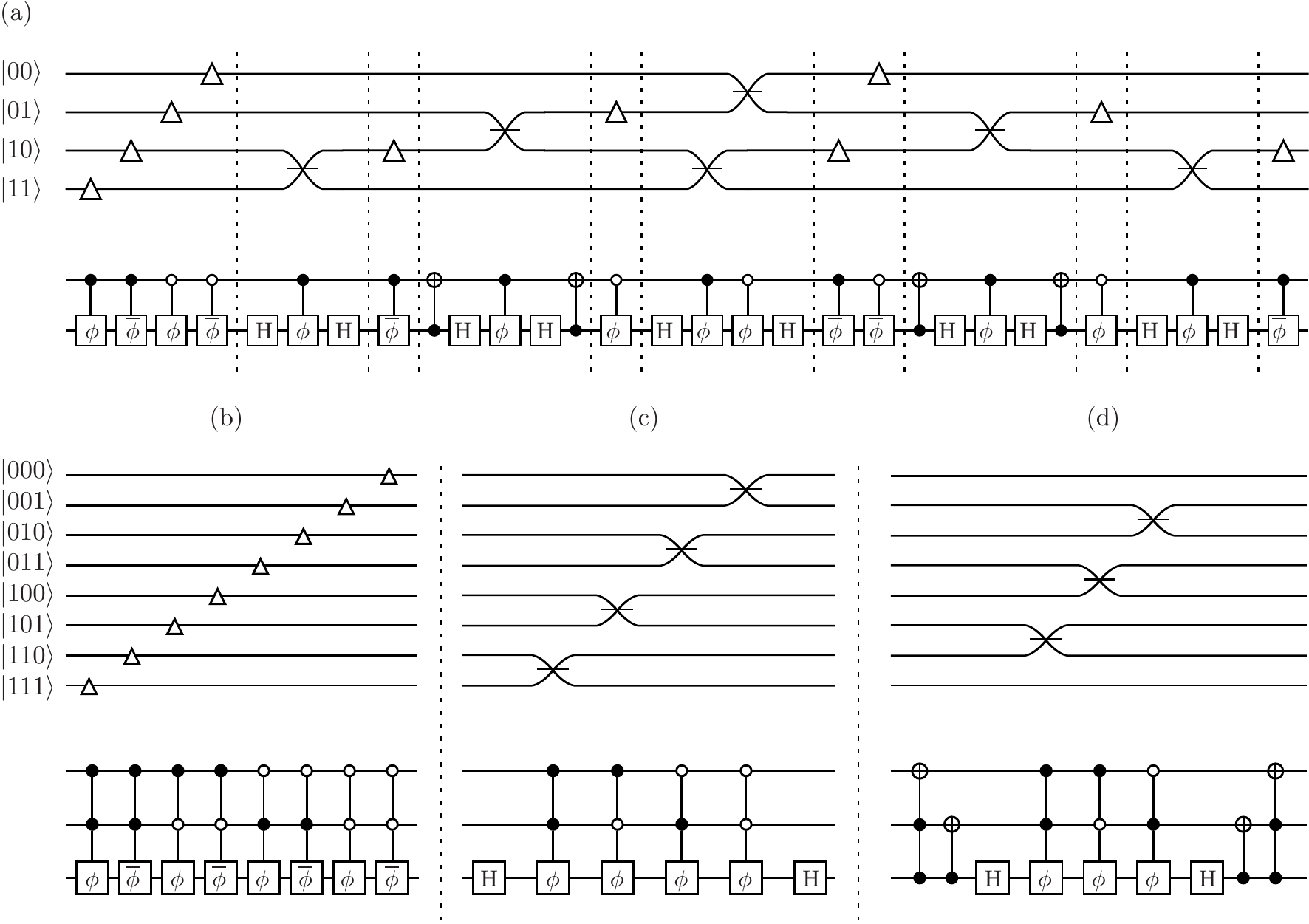}
  \caption{
Mapping a linear optical circuit to a unitary on qubits.
An empty or solid circle indicates that the operation is conditional on the `0' or `1' state of the control qubit, respectively.
(a) A 2-qubit and 4 optical mode example.
(b-d) The elementary operations for a 3 qubit (8 optical mode) unitary.
(b) Phase shifts on single elements in the qubit space result from a judicious choice of conditions for the control qubits.
(c) The addition of Hadamard gates allows beamsplitter operations to be implemented on elements in the qubit space that differ only in the state of the final qubit.
 (d) The further addition of $n$-qubit-NOT gates allows the mapping of beamsplitters for elements that may differ in the state of more than one qubit.}
  \label{fig:qubits}
\end{figure*}

Given the above parameterisation of Haar-random optical circuits, we now address the effects of errors, caused by imperfections in integrated photonics manufacturing. Before going into detail, we emphasize the important feature of our approach: due to the separability of the derived probability distributions, errors on a given component of the circuit do not propagate to other independently chosen parameters. In turn, a major source of individual errors is the imperfection of directional couplers. Used to implement the balanced beamsplitters of MZIs, directional couplers should ideally couple 1/2 of the light between waveguides so that each MZI can achieve the full reflectivity range. Fabrication tolerances, however, introduce errors and limitations on this range. Furthermore, we note that upper MZIs in the triangular scheme and central MZIs in the rectangular scheme are those most sensitive to errors, according to their polynomially growing pdfs.

Although schemes exist to minimise the effect of such errors and produce near perfect MZIs~\cite{Miller:2015,Obrien:2016} it is worthwhile considering the influence of many small errors over a large circuit (this simple model is also useful to the qubit picture that we develop below). As an estimate to this effect we address the range of unitary operations covered by the proposed parameterisation, which we evaluate in terms of the {\it coverage} of the unitary space (see references~\cite{Schaeff:2015,sp-jmp-53-013501} for more details),
\begin{eqnarray} \label{cover}
\mathrm{cov}_m=\frac{\prod_{n=2}^m \prod_{i=1}^{n-1}\int_{|\varepsilon|}^{1-|\varepsilon|} d r_{n,i} \mathcal{P}_{v_n}({\bf r}) }{\int_{\mathrm{U}} d \mathrm{U}}.
\end{eqnarray}

That is, $\mathrm{cov}_m$ is the ratio between the reachable and full unitary spaces, assuming that the phase shifters cover their full range $[0, 2 \pi)$. The range of MZI reflectivities, in turn, is $[|\varepsilon|, 1-|\varepsilon|]$, where $\varepsilon$ is a small random error. In figure~\ref{fig:coverage} we plot the coverage versus the circuit size $m$, which shows that for such moderate errors our parameterisation achieves high coverage rates. Since the pdfs for the triangular and rectangular schemes have been shown to be equivalent and independent, the coverage plotted in figure~\ref{fig:coverage} is valid for both.

We now briefly show how these results may be extended to the scenario of quantum information processing with qubits, independently of any particular physical implementation.
We suggest a mapping between a unitary operation on \(m=2^{p}\) optical modes and the same unitary operation on \(p\) qubits,
such that the \pdf{}s derived above can be directly applied to systems of qubits.
Labelling the optical modes as qubit basis states $\{|0...00\rangle, |0...01\rangle, |0...10\rangle, ..., |1...11\rangle\}$ we map the optical beamsplitters and phase shifters to single qubit Hadamard gates, and $n$-qubit logic gates where the state of a single target qubit is transformed depending on the states of $n-1$ control qubits.
The target qubit operations are the NOT gate or qubit-flip operator, $\sigma_{x}$, and the qubit-phase gates, $\Phi = e^{i \phi \sigma_{z}}$ and $\overline{\Phi} = \sigma_{x}\Phi\sigma_{x}$.
Each optical phase is mapped to a $n$-qubit \(\Phi\) or \( \overline{\Phi} \) logic gate,
and each optical beamsplitter is mapped as an MZI to a $n$-qubit \(\Phi\) or \( \overline{\Phi} \) logic gate between two single qubit Hadamard gates on the respective target qubit.

The mapping can be understood with reference to figure~\ref{fig:qubits}, which explicitly details the case for 3 qubits and 8 optical modes and present a full circuit example for 2 qubits and 4 optical modes.
The target for the $n$-qubit phase operations is always the final qubit;
the conditioning configuration of the control qubits determines which element in the qubit space receives the phase.
The addition of Hadamard operations on the final qubit allows the mapping of $1/2$ reflectivity beamsplitters, and therefore MZIs, that operate between pairs of optical modes that differ in labelling only by the final bit.
The further addition of $n$-qubit NOT gates 
allows MZIs to be mapped from pairs of optical modes that may differ in labelling by more than one bit.
Any subset of the MZI operations may be implemented on qubits by simply omitting controlled phases where appropriate.

While not designed to be optimal, this one-to-one mapping between $n$-qubit phase gates and optical MZIs illustrates one way in which the distributions expressed in figure~\ref{fig:example}(c) may be used to directly implement a HRU on qubits.

We have presented a recipe to directly generate HRUs in linear optics with a proof that is straightforward in comparison to previous works \cite{sp-jmp-53-013501, zy-jpa-27-4235}.
Experimental conformation of these results can make use of tomography that does not require further optical circuitry \cite{Laing:2012uw}.
The formula in its general form is applicable to boson sampling where Haar unitaries are required, and the extension to systems of qubits invites wider applications.

This work was completed shortly after the tragic death of one of the authors, Nick Russell.
Those of us who knew him are grateful for his contributions to our work and our lives, which we continue to miss.
We acknowledge support from the Engineering and Physical Sciences Research Council (EPSRC), the European Research Council (ERC), including QUCHIP (H2020-FETPROACT-3-2014: Quantum simulation), the U.S. Army Research Office (ARO) grant W911NF-14-1-0133. A.L. acknowledges support from an EPSRC early career fellowship.

\vspace{20pt}
\setcounter{equation}{0}
\renewcommand{\theequation}{A.\arabic{equation}}
\appendix
\section{Appendix: Converting a Lower Hessenberg matrix to a lower triangular matrix}
\vspace{-20pt}
We set \(\mat{J}_{k-1,0}^{k}=0\) with column operations by subtracting column \(\vec{c}_{k}\) multiplied by an appropriate scalar:
\begin{equation}
  \vec{c}_{0}^{\left(k\right)} = \vec{c}_{0}^{\left(k-1\right)} - \vec{c}_{k}
  \frac{\mat{J}_{k-1,0}^{\left(k-1\right)}}{\mat{J}_{k-1,k}}.
\end{equation}
The effect on all the other elements of \(\vec{c}_{0}\) is to remove the dependence on \(r_{k}\), which we can prove inductively.

Suppose that after \(k\) such operations, the upper \(k\) elements of
\(\vec{c}_{0}\) have been set to zero and the remaining elements have no
dependence on \(r_{l}\) for \(0 \leq l \leq k\). We can express the elements of
\(\vec{c}_{0}^{\left(k\right)}\) as:
\begin{equation}
  \label{eq:koperations}
  \mat{J}_{i,0}^{\left(k\right)} = \left\{ \begin{array}{lcl}
    r_{i+1} \prod_{l=k+1}^{i} \left( 1-r_{l} \right) & , & i \geq k \\
    0 & , & i < k
  \end{array} \right.
\end{equation}
The base case is \(k=0\), where the expression in (\ref{eq:firstcolumn})
corresponds to this general form. We now perform the \(\left(k+1\right){
\text{th}}\) operation on all non-zero rows (i.e.\ \(i \geq k+1\)):
\begin{align}
\nonumber
  \mat{J}_{i,0}^{\left(k+1\right)} &= \mat{J}_{i,0}^{\left(k\right)} - 
    \frac{\mat{J}_{k,0}^{\left(k\right)}}{\mat{J}_{k,k+1}} \mat{J}_{i,k+1} \\
    \nonumber
  &= r_{i+1}\prod_{l=k+1}^{i} \left( 1-r_{l} \right) + \\
  \nonumber
  &\phantom{=} \frac{\displaystyle r_{k+1} \prod_{
    l=k+1}^{k} \left( 1-r_{l} \right) }{\displaystyle r_{0} \prod_{l=1}^{k}
    \left( 1-r_{l} \right) } \frac{ r_{0} r_{i+1} }{ \left( 1-r_{k+1} \right) }
    \prod_{ l=1 }^{i} \left( 1-r_{l} \right) \\
    \nonumber
  &= r_{i+1} \prod_{l=k+1}^{i} \left( 1-r_{l} \right) + \frac{ r_{k+1} r_{i+1}
    }{ \left( 1-r_{k+1} \right) } \prod_{l=k+1}^{i} \left( 1-r_{l} \right) \\
    \nonumber
  &= r_{i+1} \left( 1-r_{k+1} \right) \prod_{l=k+2}^{i} \left( 1-r_{l} \right)
    + \\
    \nonumber
  &\phantom{=} r_{i+1} r_{k+1} \prod_{l=k+2}^{i} \left( 1-r_{l} \right) \\
  \nonumber
  &= r_{i+1} \left( 1-r_{k+1}+r_{k+1} \right) \prod_{l=k+2}^{i} \left( 1-r_{l}
    \right) \\
    \nonumber
  &= r_{i+1} \prod_{l=k+2}^{i} \left( 1-r_{l} \right).
\end{align}
We recover the expression in (\ref{eq:koperations}), thus proving the result.
After \(n-1\) iterations, we find that \(\mat{J}_{n-1,0}^{n-1}=r_{n}=1\),
recalling that \(r_{n}=1\) was a variable introduced for convenience.

\end{document}